\pdfoutput=1
\documentclass[aip,pop,amsmath,showpacs,amssymb,preprint]{revtex4-1}
\usepackage[T1]{fontenc}
\usepackage[latin1]{inputenc}
\usepackage[english]{babel}
\usepackage{graphicx}
\usepackage{bm}
\usepackage{amsmath}
\usepackage{amsfonts}
\usepackage{amssymb}
\usepackage{epsf}
\usepackage{epsfig}
\usepackage{subfigure}
\usepackage{mathrsfs}
\usepackage{ulem}

\begin{document}
\renewcommand{\v}[1]{\boldmath{#1}}
\newcommand{\ici}{images}

\title{GeV electrons due to a transition from laser wakefield acceleration to plasma wakefield acceleration}

\author{P. E. Masson-Laborde}
\email{paul-edouard.masson-laborde@cea.fr}
\affiliation{CEA, DAM, DIF, F-91297 Arpajon Cedex, France}

\author{M. Z. Mo}
\thanks{Equivalent First Author}
\affiliation{Department of Electrical and Computer Engineering,University of Alberta, Edmonton, AB, Canada,T6G 2V4}

\author{A. Ali}
\affiliation{Department of Electrical and Computer Engineering,University of Alberta, Edmonton, AB, Canada,T6G 2V4}

\author{S. Fourmaux}
\affiliation{INRS-EMT, Universit\'{e} du Qu\'{e}bec, 1650 Lionel Boulet, Varennes, Qu\'{e}bec, Canada, J3X 1S2}

\author{P. Lassonde}
\affiliation{INRS-EMT, Universit\'{e} du Qu\'{e}bec, 1650 Lionel Boulet, Varennes, Qu\'{e}bec, Canada, J3X 1S2}

\author{J. C. Kieffer}
\affiliation{INRS-EMT, Universit\'{e} du Qu\'{e}bec, 1650 Lionel Boulet, Varennes, Qu\'{e}bec, Canada, J3X 1S2}

\author{W.Rozmus}
\affiliation{Theoretical Physics Institute, University of Alberta, Edmonton T6G 2E1, Alberta, Canada}

\author{D. Teychenn\'{e}}
\affiliation{CEA, DAM, DIF, F-91297 Arpajon Cedex, France}

\author{R. Fedosejevs}
\email{rfed@ece.ualberta.ca}
\affiliation{Department of Electrical and Computer Engineering,University of Alberta, Edmonton, AB, Canada,T6G 2V4}

\date{\today}
\begin{abstract}
We show through experiments that a transition from laser wakefield acceleration (LWFA) regime to a plasma wakefield acceleration (PWFA) regime can drive electrons up to energies close to the GeV level. Initially, the acceleration mechanism is dominated by the bubble created by the laser in the nonlinear regime of LWFA, leading to an injection of a large number of electrons. After propagation beyond the depletion length, leading to a depletion of the laser pulse, whose transverse ponderomotive force is not able to sustain the bubble anymore, the high energy dense bunch of electrons propagating inside bubble will drive its own wakefield by a PWFA regime. This wakefield will be able to trap and accelerate a population of electrons up to the GeV level during this second stage. Three dimensional (3D) particle-in-cell (PIC) simulations support this analysis, and confirm the scenario.
\end{abstract}

\date{\today}
\maketitle
\section{Introduction}
The pioneering work of Tajima and Dawson \cite{Tajima} in 1979, based on the fact that an ultrashort terawatt (TW) laser propagating through an underdense plasma will excite strong plasma wakes that may trap and accelerate electrons up to high energies, has led to the development of the laser wakefield accelerator (LWFA) concept. Since then, a tremendous amount of progress has been made in improving the quality and energy of the generated electron beams, with the goal to reach the GeV-level. Finally, in the past decade, prior to which time most of the experimental accelerated electrons were characterized by an exponential energy distribution \cite{Modena,Umstadter}, high quality monoenergetic electron beams were reported by many groups \cite{Mangles,Geddes,Faure,Thomas,Hafz,Kneip,Clayton,Mcguffey12,Mo13}.
Most of these experiments, were conducted in the so-called blowout regime or ``bubble'' regime, identified in many simulations and theoretical analyses before \cite{Mora,Pukhov,Tsung,Lu2,Lu1,Esarey,Kalmykov1,Kalmykov2} where the electrons are expelled radially from the beam axis by the transverse  ponderomotive force of the laser, which creates a three dimensional (3D) cavity (the ``bubble'') empty of electrons. This bubble, full of ions, is surrounded by a sheath of relativistic electrons, and some of them can be self-trapped and then accelerated to high energy leading to a monoenergetic bunch of electrons of high quality. This acceleration of the electrons by the strong electric field inside the bubble will be limited by the dephasing length, resulting in a maximum energy gain that can be estimated from the laser and plasma parameters \cite{Lu1}:
\begin{equation}\label{LuScalingLaw}
 E_{max}(GeV)\cong 1.7 (\frac{P}{100 TW})^{1/3}(\frac{10^{18}cm^{-3}}{n_{e}})^{2/3}
\end{equation}
where $P$ is laser power and $n_{e}$ is plasma density.

Electron beam drivers can also be used to generate the full expulsion of the electrons and to create the accelerating cavity. This process called the plasma wakefield accelerator (PWFA), is know for years \cite{Chen,Rosenzweig}, and is commonly used in experiments carried out at the Stanford Linear Accelerator Center (SLAC) using GeV electron beams to drive nonlinear plasma waves \cite{Muggli,Hogan,Blumenfeld}. In the PWFA regime, the ponderomotive force of the laser is replaced by the space charge force of the electron beam in the radial expulsion of the electrons. In the PWFA regime, the phase velocity of the wake is the same as that of the electron bunch, and therefore will be independent of the plasma density. Consequently, dephasing between the accelerated electrons and the driver only occurs when the accelerated bunch obtains higher energies and velocities than the driving bunch and reaches the center of the bubble. This is not the case in LWFA, where high-energy electrons can outrun the field that moves at the group velocity, depending on the plasma density. This difference is an important advantage of the PWFA regime compared to LWFA.


There are indications in a few laser wakefield acceleration experiments to date of peak electron energies above those that would be predicted by the scaling law given by Eqn.\ref{LuScalingLaw} leading to the generation of GeV class electrons under higher density conditions \cite{Hafz,Mcguffey12,Mo13}.  Such an energy enhancement was also observed in one of the earliest 3D PIC simulations of the laser wakefield process by Tsung et al. \cite{Tsung} where it was observed that a second bunch of electrons was accelerated to 0.84 GeV.  Hafz et al. \cite{Hafz} also compare their results to PIC simulations but neither Hafz et al. or Tsung et al. gave clear explanations as to the mechanisms causing the enhanced electron energies.  Recently Hidding et al. \cite{Hidding} proposed the combination of the LWFA process with the PWFA process in separate plasmas to create and then accelerate quasi-monoenergetic electron bunches, carrying out PIC simulations indicating that a secondary 10 pC bunch of 500 MeV electrons could be accelerated up to 1 GeV by a 100 pC primary bunch of 500 MeV electrons.  Pae et al. \cite{Pae} proposed that there can be a mode transition from the LWFA process to the PWFA process within a single interaction plasma, demonstrating in a 3D PIC simulation the acceleration of a 16 pC secondary bunch of electrons up to 320 MeV by a 200 pC primary bunch of electrons with peak energy of 380 MeV.  In this case no energy enhancement was demonstrated.  In a more recent experiment \cite{Li14} a step density gas jet was employed to obtain injection at $7.5 \times 10^{18}$ $cm^{-3}$  and then acceleration over 6 mm of plasma at $3.5 \times 10^{18}$ $cm^{-3}$ producing a continuum of accelerated electrons up to 1.5 GeV in energy.  Analysis of the results using 2D PIC simulations indicated that a secondary bunch of electrons was accelerated with peak energies up to 1.8 GeV in a mechanism they describe as phase locking with the plasma wake.  However, there was no discussion as to why this phase locking occurs and no indication of a quasi-monoenergetic bunch.

In this paper, we will report the experimental results on the laser wakefield driven electron generation achieved with the 200 TW beamline at the Advanced Laser Light Source (ALLS) facility located at INRS, Varennes \cite{Fourmaux}, which is a platform developed for high intensity relativistic laser-plasma interaction studies and for laser wakefield acceleration studies \cite{ChenZL, Mo13, Mo12,Fourmaux2}. In the experiment, GeV electrons have been observed with self-injection in relatively high density plasma, on the order of $1 \times 10^{19}$ $cm^{-3}$, produced by ultra-intense laser pulse interacting with a single-stage gas jet. We will show from 3D particle-in-cell (PIC) simulations, that the level of energy obtained and the characteristics of the electron beams can be understood as a two-stage process, where in the first stage the LWFA will accelerate a dense bunch of high energy electrons, and then after the complete depletion of the laser pulse, this bunch will create a wakefield in the PWFA regime, which is able to accelerate electrons close to GeV level. The PIC simulation analysis implies that these GeV electrons can be seen as experimental observation of the two-stage process describe by Ref.[\onlinecite{Hidding}], indicating that the single-stage hybrid plasma wakefield acceleration is a feasible approach to achieve energetic electron beams.

The paper is organized as follows. Section \ref{sec2a} and \ref{sec2b} present the experimental setup and the experimental results respectively. Section \ref{sec3} presents the simulation results of the experiment and their interpretation. Section \ref{sec4} and \ref{sec5} summarize our discussions and conclusions respectively.

\section{Experimental Setup}\label{sec2a}

\begin{figure}[!h]
\resizebox{13.cm}{!}{\includegraphics{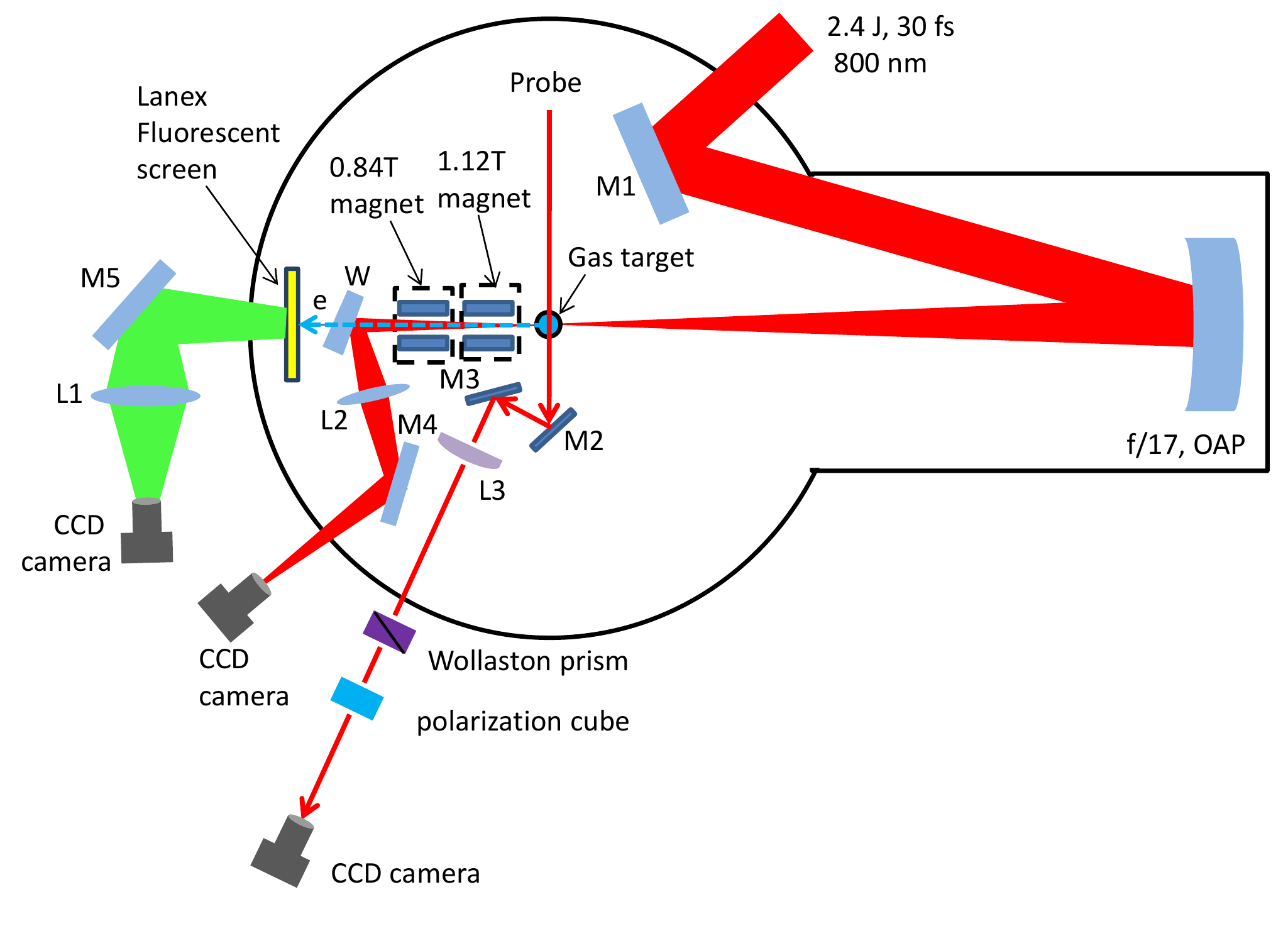}}
\caption{Schematic diagram of the experimental setup for the laser wakefield acceleration experiments. M1 - M5: reflective mirrors; OAP: off-axis parabola; L1 - L3: lenses; W - beam splitter glass wedge.}
\label{ExptSetup}
\end{figure}

The experiments were performed with the 200 TW laser system located at the Canadian Advanced Laser Light Source(ALLS) facility at INRS, Varennes.\cite{Fourmaux} The 200 TW laser system is a compact laser system based on Ti:Sapphire technology and chirped pulse amplification (CPA) technique with a central operating wavelength of 800 nm in horizontal polarization. For typical data shots during the experiments, the facility delivered laser pulses with energy of 2.4 J and pulse duration of 30 fs at full-width half-maximum (FWHM) onto the gas target. As shown in Fig.\ref{ExptSetup}, the 9-cm-diameter laser pulses were focused by a 150 cm focal length off-axis parabola (OAP) onto the gas target. The vacuum focal spot measured with a single-lens imaging system shows a full-width at half maximum (FWHM) diameter of approximately 22 $\mu m$, within which area it contains $\sim25\%$ of the total energy. The focused peak intensity in vacuum was measured to approximately $7.0 \times 10^{18}$ $W/cm^{2}$, corresponding to a laser normalized vector $a_{0}$ of 1.7.

The generated electron beams were dispersed by two separate 10-cm-long dipole magnets with magnetic field strengths of 1.12 T and 0.84 T onto a Lanex fluorescent screen that was placed 20 cm after the last magnet. The fluorescence emitted from the Lanex screen was collected by an f/2.8 aperture lens system and imaged onto a 12-bit charge coupled device(CCD) camera. A side-view Normarski interferometer based on a Wollaston prism as the beam splitter \cite{Benattar79} was employed to monitor the plasma density. The probe beam for the side-view interferometry came from the zero-order diffraction of the first grating in the compressor, which was then compressed with an extra compressor down to 40 fs. The path length of the probe beam is adjustable to get various delays relative to the main pulse. The Thomson scattering light emitted from the laser plasma interaction region at an angle normal to the plane spanned by the laser polarization direction and the propagation direction was collected by a top-view imaging system to monitor the plasma channel formation.

The gas jet was formed by a 5-mm-diameter supersonic conical nozzle connected to a pulsed solenoid valve (Parker Valve). The working gas during experiments is pure helium. The density of the helium plasma was calculated by use of modified Abel inversion algorithm\cite{Fed79}, where the asymmetry of the fringe shifts is weighted and introduced into the final plasma distribution, assuming a $\cos\theta$ transverse asymmetry contribution. The uncertainty of the measured electron energy was estimated according to an electron beam shot to shot divergence of 9.8 mrad, which is derived based on the standard deviation of the positions of the straight through reference shots when both magnets were removed, leading to an error of (+311 MeV$\backslash$-196 MeV) at 1 GeV.


\section{Experimental Results}\label{sec2b}


\begin{figure}[!h]
\resizebox{13.cm}{!}{\includegraphics{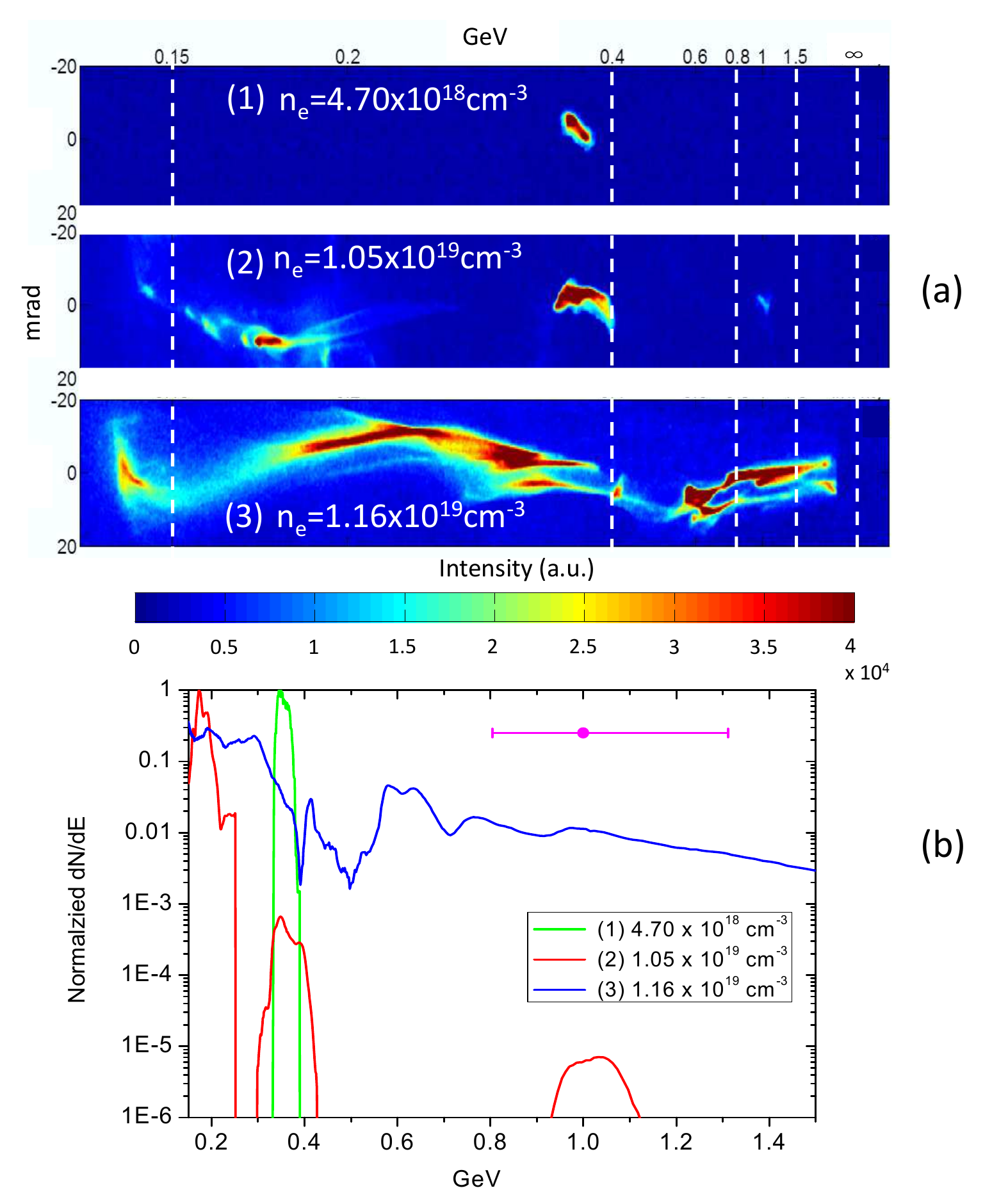}}
\caption{(a) Energy resolved images of the electron bunches for pure helium at plasma densities, (1) $4.70 \times 10^{18}$ $cm^{-3}$, (2) $1.05 \times 10^{19}$ $cm^{-3}$, (3) $1.16 \times 10^{19}$ $cm^{-3}$;  All the images are plotted in the same color range where the brightness represents the flux of the electrons in arbitrary units. (b) Corresponding normalized electron number density per electron energy; Note that the y axis is in logarithmic scale; Representative uncertainty of measured electron energy at 1 GeV is indicated by the magenta circular dot and attached bars at the top of the plot.}
\label{Espectra}
\end{figure}

\begin{figure}[!h]
\resizebox{13.cm}{!}{\includegraphics{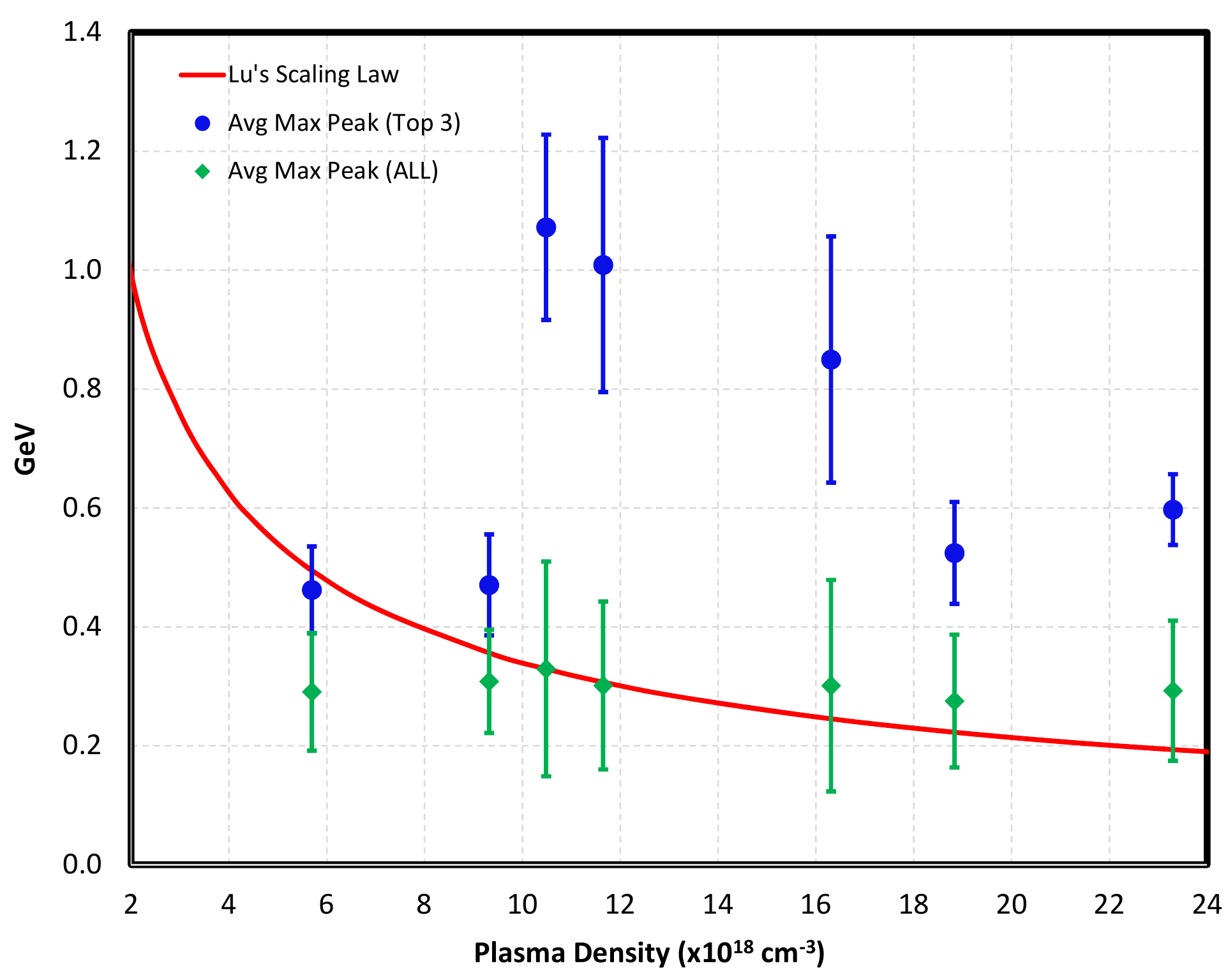}}
\caption{Energy of the highest energy peak in the electron distribution measured at each electron density for pure helium. The blue dots stand for the average of the top three maximum achieved peak energies at each density. The green diamonds are the averages of all the energies of the highest energy peaks for shots at identical density. The error bars are obtained from the standard deviation of the given number of measurements. The red line represents the predicted energies at a given laser power of 80 TW using the nonlinear scaling law given by Eqn.[\ref{LuScalingLaw}]}
 \label{PEnery}
\end{figure}

\begin{figure}[!h]
\resizebox{13.cm}{!}{\includegraphics{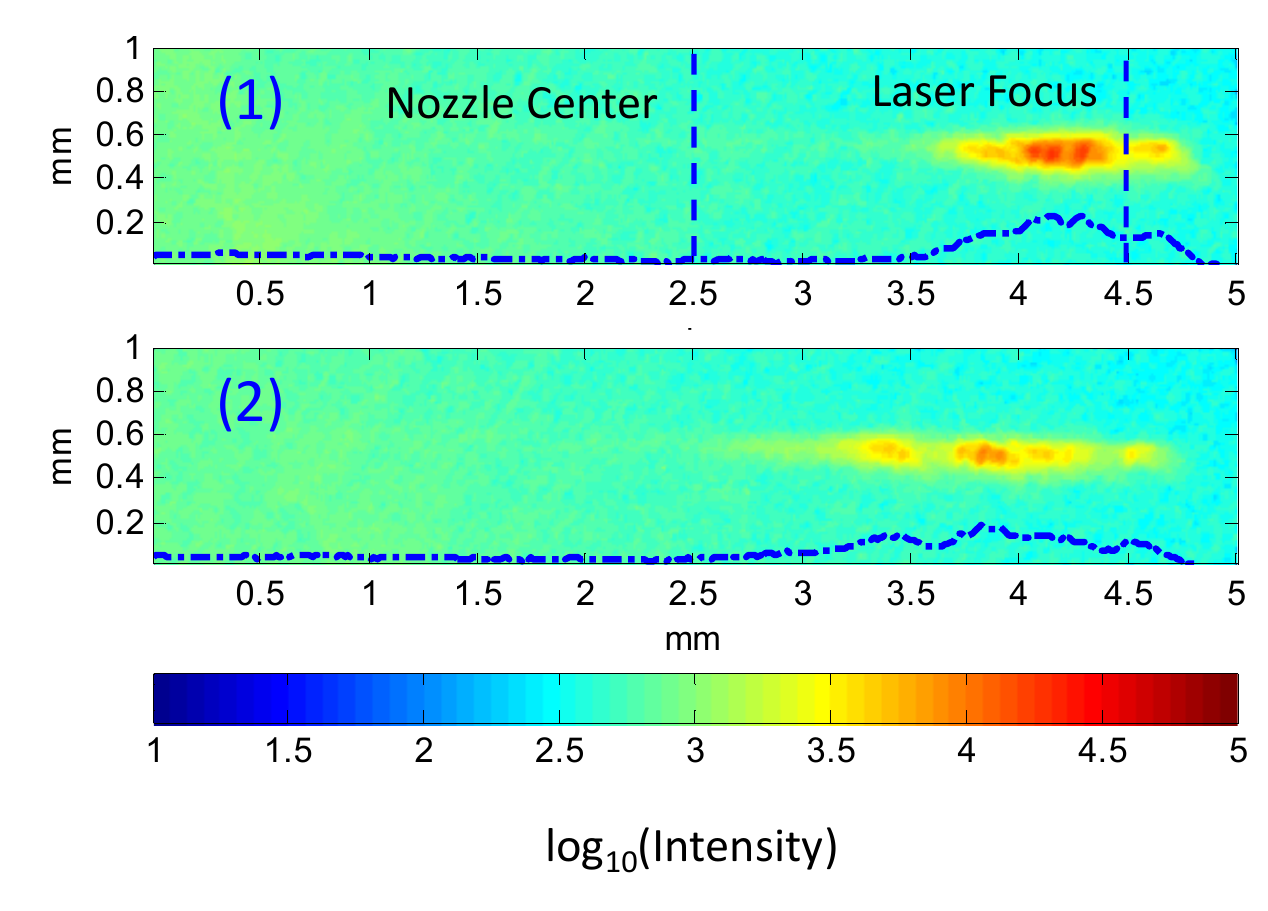}}
\caption{(1)-(2) Thomson scattering images of the plasma channels formed at identical density of $1.16 \times 10^{19}$ $cm^{-3}$ achieved with the top-view imaging system. Note that the false color in all of the images is plotted in base 10 logarithmic scale to present clearer features of the plasma channel. Lineouts of the plasma channels after taking logarithm are rescaled and overlaid with the images. The nozzle center and the laser focus position are marked with blue dashed lines. The laser propagated from right to left.}
\label{Topview}
\end{figure}

The electron energy spectra obtained at different plasma densities: $4.70 \times 10^{18}$ $cm^{-3}$, $1.05 \times 10^{19}$ $cm^{-3}$ and $1.16 \times 10^{19}$ $cm^{-3}$ using 5-mm gas jet with laser power of 80 TW are shown in Fig.\ref{Espectra}(a). The electron energy is dispersed in the horizontal direction while the vertical profile shows the lateral deflection and divergence of the electron beams. The corresponding $dN/dE$ electron energy spectra integrated over the full width for each spectral image normalized to unity is plotted in Fig.\ref{Espectra}(b). Pronounced monoenergetic peaks ranging from 0.15 to over 1 GeV have been observed in different shots as shown in Fig.\ref{Espectra}(a). More distinct monoenergetic features were observed at the lower electron density of $4.70 \times 10^{18}$ $cm^{-3}$, while multiple bunches and quasi-continuous injection start to dominate as the electron density was increased to $1.05 \times 10^{19}$ $cm^{-3}$ and $1.16 \times 10^{19}$ $cm^{-3}$.

The top image of Fig.\ref{Espectra}(a) shows a typical electron image at plasma density of $4.70 \times 10^{18}$ $cm^{-3}$. From the image, one can see that there is merely one monoenergentic electron bunch in the spectrum, which has a prominent peak at 345 MeV with energy spread of $10\%$, a total charge of 7.3 pC and $1/e^{2}$ beam divergence of 12 mrad. When increasing the plasma density to $1.05 \times 10^{19}$ $cm^{-3}$, as shown in the middle image of Fig.\ref{Espectra}(a), three separated electron bunches of different characteristics are observed. To the high energy end, a relatively week monoenergetic electron bunch, having a total charge of 0.4 pC, peaks around $1.03$ GeV. The $1/e^{2}$ beam divergence of this GeV bunch was measured to be less than 6 mrad. The second bunch to the left, peaking around 350 MeV, is also monoenergetic and has a total charge of 11.7 pC. Comparing the first two electron bunches, the third one peaking around  175 MeV spreads more in divergence but contains a larger total charge of 36.6 pC. At an electron density of $1.16 \times 10^{19} cm^{-3}$, as shown in the bottom image of Fig.\ref{Espectra}(a), more continuous injection was observed with high energy electrons extending up to $1.5^{+0.8}_{-0.4}$ GeV.  These features can clearly be seen in the lineout plots of electron number density per unit energy in Fig.\ref{Espectra}(b). The total charge contained in this shot is measured to be around 245 pC. Note that the calculated charge for these electron images is based on the manufacturer's specifications for the camera response and the imaging system optical efficiency. It is estimated that the accuracy of this result is within a factor of 2.

Comparing the maximum peak energies achieved for these three particular shots, one may notice that higher energy bunches of electrons appear at the higher densities which is contrary to the expected decreasing dephasing length and lower maximum energy obtained with increasing density.  Also, applying Lu's scaling law for maximum energy gain as shown in Eqn.\ref{LuScalingLaw}, one would expect a peak electron energy of 350 MeV at the plasma density around $1.1 \times 10^{19}$ $cm^{-3}$ for a laser power of 80 TW as employed in this experiment. However, the measured GeV peak energies around this density as depicted in the bottom two images in Fig.\ref{Espectra}(a) are more than double of the expected value, which as we will demonstrate using PIC simulations is due to some energy boost mechanism introduced by the transition from LWFA to PWLA occurring at high plasma density. The details will be discussed in Sec.\ref{sec3}. Energy spectra shown in Fig. \ref{Espectra} illustrate transition from the LWFA mechanism to PWFA with increasing plasma density. As the PWFA becomes more efficient at higher background densities the energy spread of accelerated electrons also becomes larger consistent with a theory of PWFA \cite{Esarey}.

For each shot the maximum energy peak was determined where a distinct peak in the energy spectrum was observed. This is called the maximum peak energy. Fig.\ref{PEnery} plots the average of the maximum peak energies achieved at different plasma densities. The green diamonds represent the average of all the maximum peak energies achieved at identical plasma density. An average of 60 shots at each density were taken to conduct the statistics of the maximum peak energies. The blue dots stand for the average of the top three maximum peak energies for each density. For comparison, the predicted average maximum energy gain at each density according to Eqn.\ref{LuScalingLaw} with input laser power of 80 TW is also plotted in the graph.  Looking at the average of all the maximum peak energies, one can see that the peak energy agrees approximately with the prediction when the plasma density is above $6 \times 10^{18}$ $cm^{-3}$, below which the measurement starts to deviate from the prediction. This phenomena was observed and reported previously \cite{Mo12} and is attributed to the violation of bubble matching condition which is needed to self guide the laser propagation inside the plasma. However, in relatively high plasma density region, i.e. around $1\times 10^{19}$ $cm^{-3}$ and above, occasionally there are some shots that generate electrons more than double of the prediction, particularly in the region near $1.1\times 10^{19}$ $cm^{-3}$, in which GeV electrons were observed. A 3D Particle-in-cell (PIC) simulation was conducted to understand the physics behind this energy doubling phenomena and the details will be given in Sec.\ref{sec3}.

The laser propagation in the plasma can be experimentally studied by looking at the Thomson scattering light emitted from the plasma.\cite{Gibbon96} To aid in the understanding of the physics behind the observed energy boosting phenomena, the plasma channel images were captured during the experiments and are shown in Fig.\ref{Topview}, which plots two typical Thomson scattering images achieved at the same plasma density of $1.16 \times 10^{19}$ $cm^{-3}$ for pure He. Among them, the top one is the plasma channel image corresponding to the GeV shot (the bottom image) as shown in Fig.\ref{Espectra}(a). As indicated, the plasma channels appear around 0.25 mm before the laser focus and last for less than 2 mm due to the pump depletion. Particularly, for the GeV shot achieved at this density as shown in the Fig.\ref{Espectra}(a), the plasma channel is around 1.3 mm in length. At the density of $1.16 \times 10^{19} ~cm^{-3}$, the plasma wavelength is estimated to be $9.82 ~\mu m$ by using the formula: $\lambda_{p}[\mu m]=3.34 \times 10^{10} / \sqrt{n_{e}(cm^{-3})}$, and the pump depletion length, approximated by \cite{Lu1}$L_{pd}=c\tau_L(\omega_0/\omega_p)^2$, where c is the light speed in vacuum and $\tau_L$ is the laser pulse duration, $\omega_0$ and $\omega_p$ are the laser and plasma angular frequencies, is estimated to approximately 1.4 mm. The estimated pump depletion length agrees approximately with the length of plasma channel that we observed here. The critical power for self-focusing, given by $P_{c}(GW)=17 n_{c}/n_{e}$, where $n_{c}$ is the critical density given by $n_{c}=\omega^{2}m\varepsilon_{0}/e^{2}$ ($\varepsilon_{0}$ is the permittivity in free space), is calculated to be $\sim 2.6$ TW at the density of $1.16 \times 10^{19} ~cm^{-3}$, which gives a ratio of $P/P_{c}$ of 31 for 80 TW laser. With the laser power greatly in excess of the critical power for self-focusing, the modulated structures of the Thomson images, as shown in Fig.\ref{Topview}, come about as a general response to the self-oscillations of beam propagating in a nonlinear medium.\cite{Gibbon96}

\section{Simulations results}\label{sec3}

To understand the physics behind the energy doubling phenomena that we observed, we use the 3D fully relativistic PIC code SCPIC \cite{Popov}, which is a successor of the code MANDOR \cite{Romanov,Mordovanakis} already used in many laser-plasma applications. This code uses a well-known Yee scheme for solving Maxwell equations and the Boris scheme for equations of motion of the macroparticles. The laser pulse propagates in the x-direction along a fully ionized plasma (y and z being transverse directions). Parameters are the same as those used in the experiment, the laser wavelength is $\lambda_0=0.8$ $\mu m$, the pulse duration is $\tau_L=30$ fs at full-width-half-maximum (FWHM), and the laser focal spot is $w_0=22$ $\mu m$ at FWHM. The laser pulse peak intensity is $I=7\times10^{18}$ $W/cm^2$, corresponding to a normalized vector potential $a_0=1.7$. The laser is propagating in a 5-mm-long plasma, with a homogeneous density of $n_e=1.1\times10^{19}$ $cm^{-3}$, corresponding to a normalized density of $n_e/n_c=5.7\times10^{-3}$ with $n_{c}$ the laser critical density. In our study, we use a moving window and the simulation window size is $125\lambda_0 \times 78\lambda_0 \times  78\lambda_0$, and the total number of macroparticles used is about $1.2\times10^{8}$.
Based on our experimental conditions, the pump depletion length \cite{Lu1} is $L_{pd}=c\tau_L(\omega_0/\omega_p)^2\approx1.43$ $mm$, and the dephasing length is $L_d=(2/3)\sqrt{a_0}(\omega_0/\omega_p)^2\lambda_p/\pi\approx0.9$ $mm$, both are much smaller than the total length of the plasma, so we may expect that the laser will not be able to maintain sufficient power to accelerate electrons up to the total 5 mm long. Note that the estimate of the pump depletion length, $L_{pd}$, is consistent with the Thomson scattering data shown in Fig.\ref{Topview} where the strong scattered light signal extends over the distance comparable with $L_{pd}$. We will show that, in agreement with Ref.[\onlinecite{Pae}], two regimes can be identified in our simulations: an early stage takes place when $t < t_{pd} \approx L_{pd}/c \approx 5$ $ps$, during which period the main acceleration mechanism is attributed to LWFA, and the second stage with mechanism due to PWFA comes into play when $t\gg t_{pd}$. As we will discuss later, the transition from the LWFA to PWFA occurs because the laser is fully depleted and cannot sustain the bubble anymore.

\subsection{LWFA regime}\label{sec3a}

\begin{figure}[!h]
\resizebox{13.cm}{!}{\includegraphics{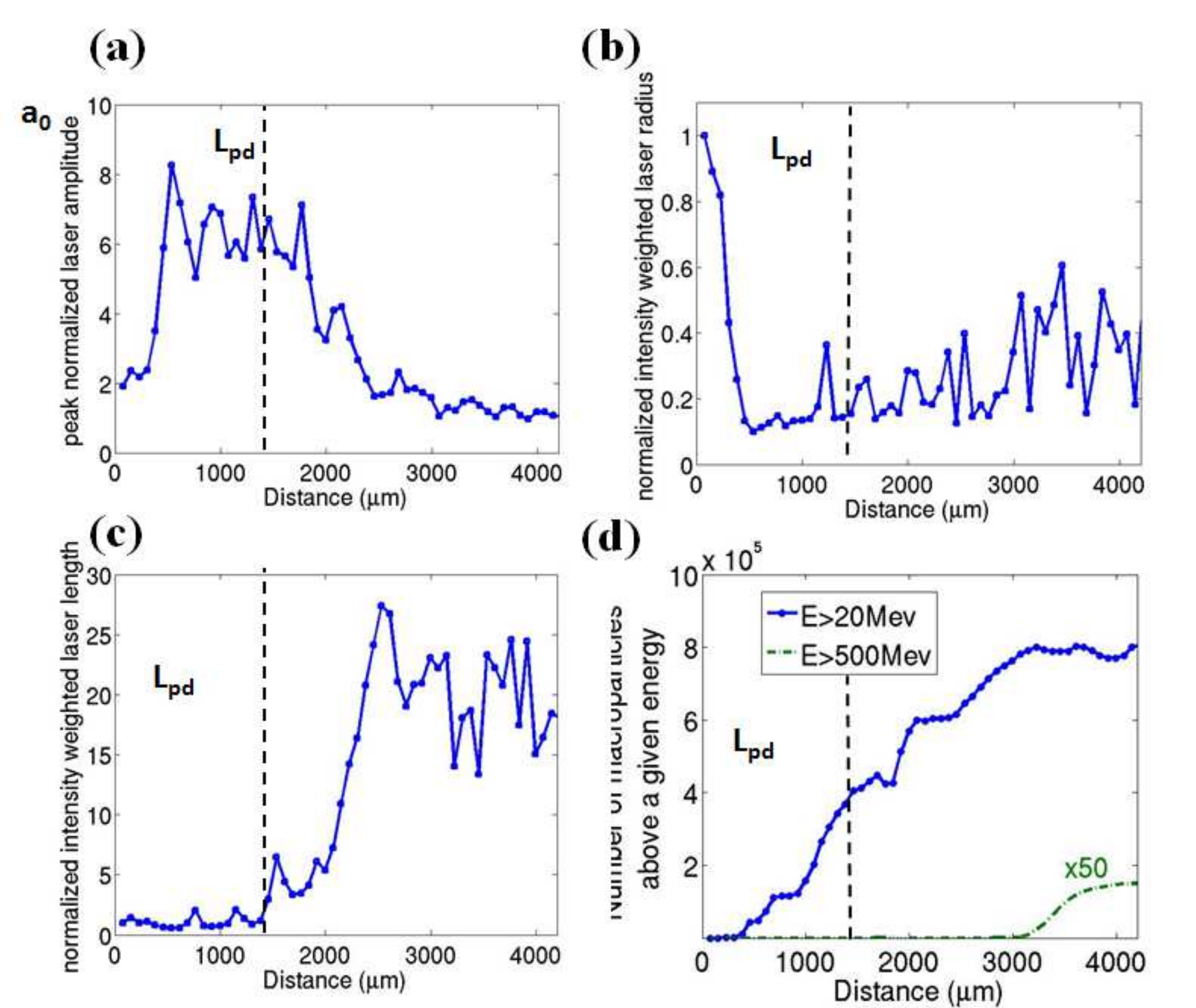}}
\caption{(a) Normalized laser peak amplitude as a function of propagation, (b) normalized intensity weighted laser radius as a function of propagation, (c) normalized intensity weighted laser length as a function of propagation, (d) number of macroparticles with energy larger than 20MeV (solid line) and larger than 500MeV as a function of propagation (dashed line, multiply by 50 in order to fit in linear scale with the 20MeV curve). The pump depletion length $L_{pd}$ is illustrated in all curves with the dashed line.}
 \label{Fig5}
\end{figure}

\begin{figure}[!h]
\resizebox{13.cm}{!}{\includegraphics{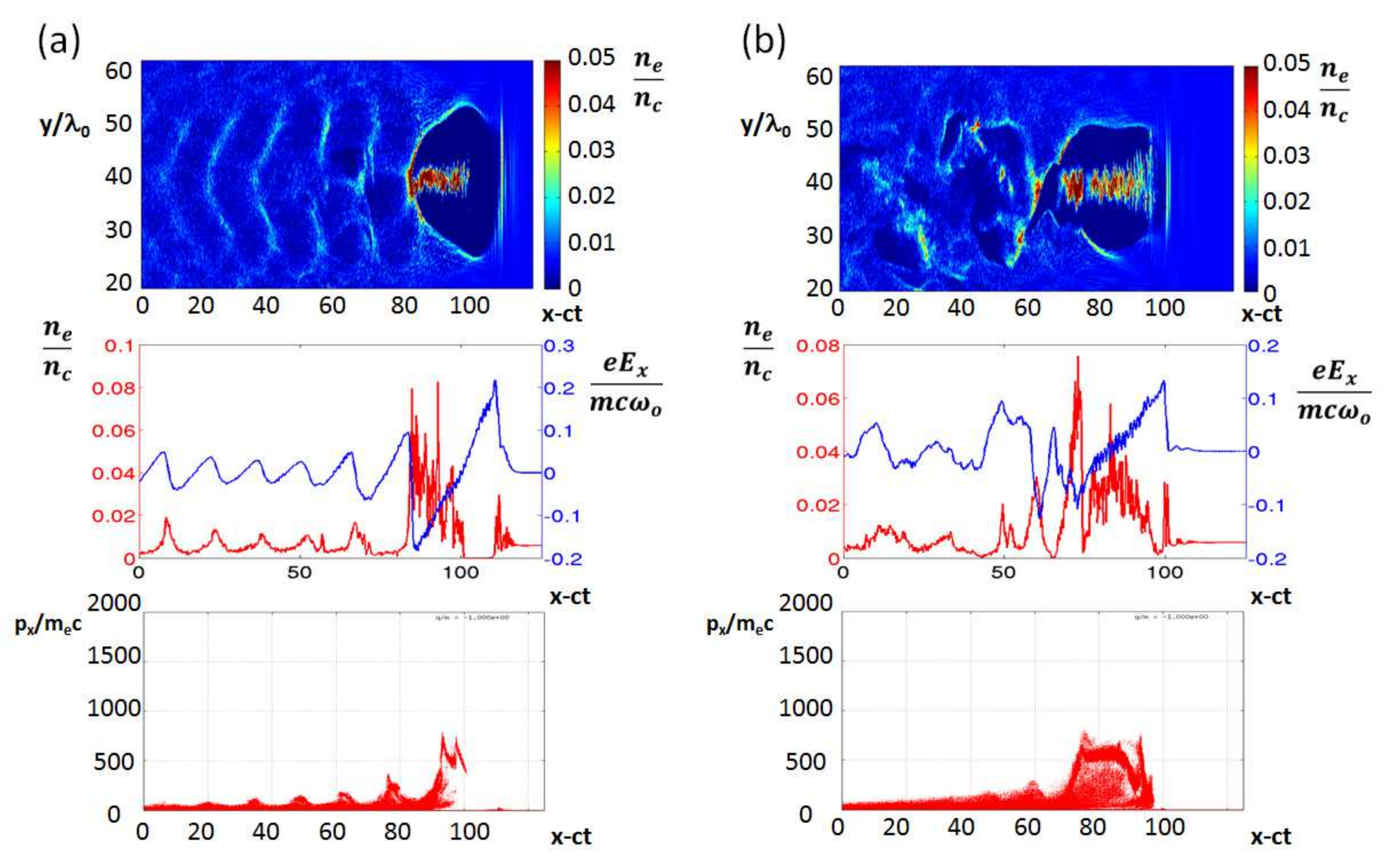}}
\caption{(color online) (a) At time t=4 ps, corresponding to a propagation distance of 1.2 mm (LWFA regime): (from top to bottom) 2D map (x-y plane) of electron density (normalized to $\rho_c$); On-axis lineout of the longitudinal electric field (blue curve) and electron density (black curve); Electron normalized momentum $p_z/mc$. (b) Same pictures but for time t=7.5 ps, or a propagation distance of 2.2 mm (end of LWFA regime). Both 2D maps are scaled to $n=0.05n_c$}
 \label{Fig6}
\end{figure}

In the first stage of the propagation, before the pump depletion length, the laser will undergo strong self-focusing (here we have $P/P_c\simeq30$), and consequently the normalized peak amplitude $a_0$ will grow up to large value, reaching a maximum around $a_0\simeq7$ (Fig.\ref{Fig5}(a)). At the same time, the radius of the focal spot will be significantly reduced. Many useful diagnostics of the pulse evolution can be found in Ref.[\onlinecite{Upadhyay}], and some of them were employed here to study the pulse evolution during the propagation. As can be seen in Fig.\ref{Fig5}(b), which shows the normalized intensity weighted laser radius given by: $<w^2>=\frac{\int y^2E_y^2dy}{\int E_y^2dy}$ ,\cite{Upadhyay} the laser radius decreases to a very small value as its amplitude is growing (Fig.\ref{Fig5}(a)) in the first 500 um's of propagation.
The transverse ponderomotive force of the laser expels the electrons and creates an ion channel. When the Coulomb force from the ion channel, which tends to pull back the expelled electrons, is equal to the radial ponderomotive force, a stable bubble shape is reached. From this equality, we can estimate the blowout radius given by $R\simeq2\sqrt{a_0}/k_p\simeq10\lambda_0$, where $k_p$ is the plasma wavenumber, given by $k_p=(2\pi/\lambda_0)\sqrt{n_e/n_c}\simeq0.6$ $\mu m^{-1}$. Fig.\ref{Fig6}(a) illustrates, after a propagation distance of $L=1.2$ $mm$: (i) the electron density $n_e/n_c$  in 2D plane (x-y, z=0) (top picture), (ii) lineout of electron density and longitudinal electric field $E_x$ (middle picture) and (iii) electron normalized momentum $p_x/m_ec$ (bottom picture). As indicated, electrons are trapped in this cavity and are accelerated up to $250$ $MeV$, and the total charge for these electrons is around 300 pC. The laser continues to propagate and electrons are injected and accelerated as long as the bubble still exists. However, after a propagation of $L=2.2$ $mm$, as seen in Fig.\ref{Fig6}(b), the bubble is elongated in the longitudinal direction due to the injection of a large amount of electrons, along with the fact that electrons injected inside the cavity start to dephase, despite the fact that the accelerating field is not yet affected too much by the beam loading \cite{Tzoufras,Vafaei} and keeps a sharp profile along the longitudinal direction. The onset of the electron injection during LWFA regime, is related to the succession of self-focusing and defocusing periods during the laser propagation \cite{Oguchi}. As depicted in Fig.\ref{Fig5}(b) and (d), which show the laser radius and the number of macroparticles above 20 MeV as a function of propagation distance respectively, the injection and acceleration starts when the minimum laser radius is reached, at which time the laser becomes self-guided.

\begin{figure}[!h]
\resizebox{13.cm}{!}{\includegraphics{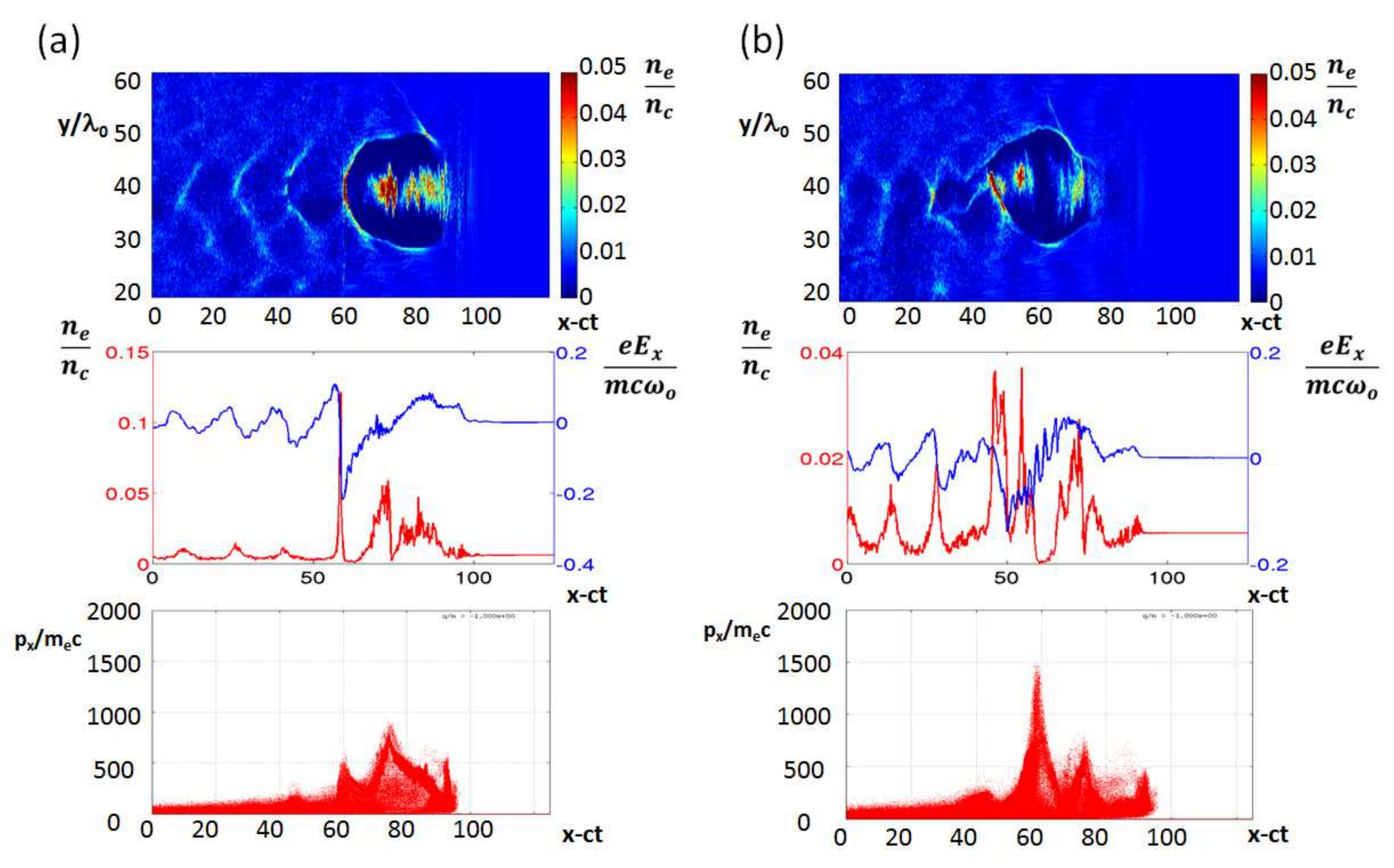}}
\caption{(color online) (a) For time t=9 ps, at a propagation distance of L=2.7 mm (beginning of the PWFA regime): top: 2D map (longitudinal-transverse plan) of electron density (normalized to $\rho_c$), middle: on-axis line out of the longitudinal electric field (blue curve) and electron density (black curve) and bottom: electron normalized momentum $p_z/mc$. (b) same pictures but for time t=12 ps, so a propagation distance of L=3.6 mm (PWFA regime). All the 2D map are scaled to $n=0.05n_c$.}
 \label{Fig7}
\end{figure}

\subsection{Transition from LWFA regime to PWFA}\label{sec3b}

As we can see in Fig.\ref{Fig5}(a), beyond the distance of laser pump deletion, the laser pulse peak amplitude begins to strongly decrease from its maximum value obtained during self-focusing. For a propagation distance of $L = 2.2$ $mm$ (corresponding to Fig.\ref{Fig6}(b)), the peak amplitude is back to its initial value. After this distance of propagation, the shape of the pulse is strongly modified from its initial shape due to group velocity dispersion and self-steepening \cite{Sprangle,Shadwick,Decker1,Decker2,Decker3}. Again, following Ref. [\onlinecite{Upadhyay}], we can look at the normalized pulse length given by: $<L_x^2>=\frac{\int (x-x_{max})^2E_y^2dx}{\int E_y^2dx}$, where $x_{max}$ is the position (longitudinal) of the maximum amplitude of the laser pulse. This quantity is illustrated in Fig.\ref{Fig5}(c), and we can clearly see that after 2 mm of propagation the depletion of the pulse occurs, leading to a broadening of the pulse length.

\begin{figure}[!h]
\resizebox{12.cm}{!}{\includegraphics{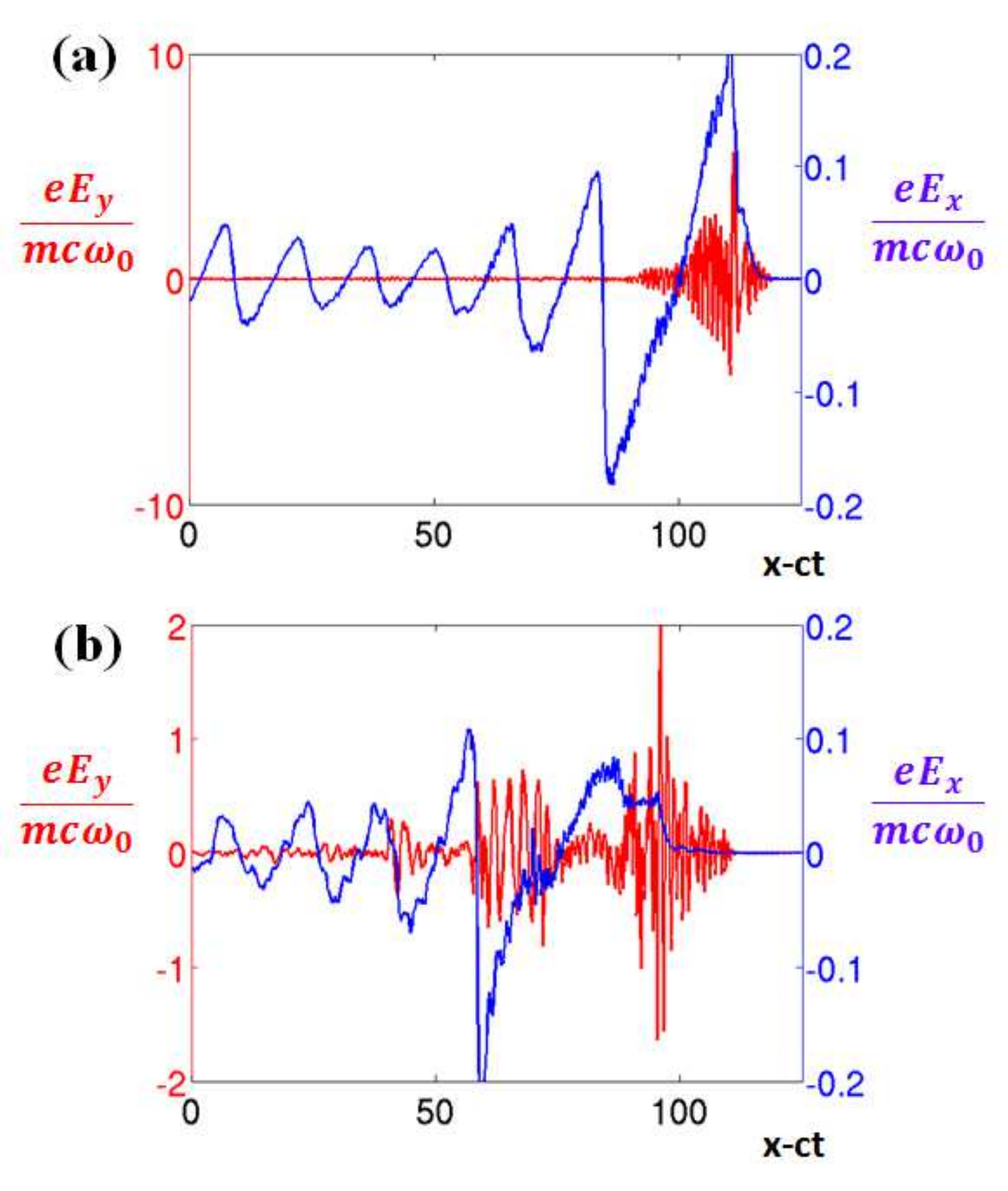}}
\caption{On-axis lineout of the longitudinal electric field (black curve) and transverse electric field (blue curve) as a function of the propagation distance for: (a) t=4 ps, at a propagation distance of L=1.2 mm (top picture) and (b) for time t=9 ps, at a propagation distance of L=2.7 mm.}
 \label{Fig8}
\end{figure}

From this moment, as can be seen in Fig.\ref{Fig7}(a) showing density, longitudinal electric field and phase space after a propagation distance of L = 2.7 mm, the electrons are dephased, and despite the low quality of the transverse electric field, the bubble is maintained. However, the longitudinal electric field does not have a sharp profile any more but also is affected by a beam loading effect. The loaded charge $Q$ and the accelerating field modified by $Q$ satisfy the following relation \cite{Tzoufras}:
\begin{equation}\label{beamloadingthresh}
 Q(nC)> 1.5\times10^{-3} \sqrt{n_e/n_c}(k_pR_b)^4 ({\frac{eE_x}{mc\omega_0}})^{-1},
\end{equation}
which illustrates the balance between the number of accelerated electrons and the field, $E_x$, produced by the original bunch. This is characteristic for the PWFA and the interaction between loaded charge and the accelerating bunch \cite{Tzoufras}. Here, for our parameters, we find that beam loading will become significant for $Q>1nC$, which is very close to the total charge of the electrons injected inside the cavity up to this propagation distance in the LWFA regime. The shape of the longitudinal electric field shows evidences that the first leading bunch of electrons inside the cavity, previously injected during LWFA regime, is now driving the acceleration by playing the role of the wakefield driver. Comparisons between longitudinal and transverse electric field at this time and the time corresponding to Fig.\ref{Fig6}(a), are illustrated respectively in Fig.\ref{Fig8}(b) and (a). We can see that, while initially the laser pulse is driving the wakefield, at late time, the position of the electric field indicates that the remaining pulse is no longer the driver of the wakefield. Furthermore, we can observe that, as already indicated by Fig.\ref{Fig5}, the laser pulse is fully depleted and broadened.
In order to be sure that the laser is not driving the wakefield, we made PIC simulations using as initial conditions, the laser as it is at the end of the LWFA regime. Our simulations confirm that,  because of the effective length and amplitude of the pulse, no wakefield can be generated at this time.

Another population of electrons is then being accelerated by the longitudinal electric field now driven by the bunch of electrons (position $x-ct\simeq60$ on Fig.\ref{Fig7}(a) and (b)).

\subsection{PWFA regime}\label{sec3c}

When the transition to plasma wakefield acceleration regime occurs, we can try to estimate the spatial features of the main electron bunch now driving the wakefield. By assuming a bi-Gaussian profile for the leading electron bunch such that $n_b=n_{b0}\exp{(-x^2/2\sigma_x^2)}\exp{(-r^2/2\sigma_r^2)}$, with $r^2=y^2+z^2$, we can obtain for the time corresponding to Fig.\ref{Fig7}(a), some estimates of the FWHM radius along the longitudinal axis: $\sigma_x\simeq5\lambda_0$ , such that $k_p\sigma_x\simeq2$, and also estimates along the transverse direction:  $\sigma_r\simeq1.2\lambda_0$ , such that $k_p\sigma_r\simeq0.6$. At this particular moment, when the second injection occurs boosted by the longitudinal electric field created by the bunch, the bunch density can be estimated as $n_b/n_e\simeq7$. Some values estimated at different times during propagation can be found in Table \ref{Tab1}.

\begin{table}[!h]
\caption{\label{Tab1} Estimates of FWHM radius $\sigma_x$ and $\sigma_r$ along longitudinal and transverse directions for different times, and leading electron bunch density estimates $n_{b0}/n_e$.}
\begin{tabular}{|c|c|c|c|c|}
  \hline
  \hspace{.1cm} Time (ps) \hspace{.1cm} &  \hspace{.1cm} $L$ (mm) \hspace{.1cm} & \hspace{.25cm} $k_p\sigma_x$ \hspace{.25cm} & \hspace{.25cm} $k_p\sigma_r$\hspace{.25cm}  & \hspace{.25cm} $n_{b0}/n_e$\hspace{.25cm} \\
  \hline
 \hspace{.1cm}  8    \hspace{.1cm} & \hspace{.1cm} 2.4   \hspace{.1cm} & 2.3   & 0.6  & 7  \hspace{.1cm} \\
 \hspace{.1cm}  10.5 \hspace{.1cm}  & \hspace{.1cm} 3.1   \hspace{.1cm} & 1.9   & 0.47 & 4.5 \hspace{.1cm} \\
 \hspace{.1cm} 12    \hspace{.1cm}  & \hspace{.1cm} 3.6   \hspace{.1cm} & 0.94  & 0.66 & 3  \hspace{.1cm} \\
 \hspace{.1cm} 12.5  \hspace{.1cm}  & \hspace{.1cm} 3.8   \hspace{.1cm} & 0.8   & 0.8  & 2  \hspace{.1cm} \\
  \hline
\end{tabular}
\end{table}

The electron bunch will be able to excite a nonlinear plasma wake if the bunch length is approximately the same order as the plasma period, e.g. $k_p\sigma_x\leq1$ according to the theory developped by Lu \textit{et al} \cite{Lu2}. As we can see, after more than 2.4 mm of propagation, when the laser is fully depleted, this condition is satisfied. Also, the transverse gradient in the bunch profile is always such that $k_p\sigma_r<<1$, and the bunch density is such that $1<n_{b0}/n_e<10$. According to linear theory\cite{Lu2}, which is valid for a narrow short electron beams such that $k_p\sigma_r<<1$ and $n_{b0}/n_e<10$, the accelerating field in the blow-out regime is maximized for $k_p\sigma_x\simeq \sqrt{2}$ and is given by:
\begin{equation} \label{eq:1}
\frac{eE_x}{mc\omega_0}\equiv\sqrt{\frac{n_e}{n_c}}1.3\frac{n_{b0}}{n_e}k_p^2\sigma_r^2\ln{\frac{1}{k_p\sigma_r}}
\end{equation}
Using estimates above, we can find that once we enter into the regime of PWFA, the theoretical value of maximum field is given by $\frac{eE_x}{mc\omega_0}\approx0.12$, which is close to results from PIC simulation as we can seen on Fig.\ref{Fig8}(b).

\begin{figure}[!h]
\resizebox{13.cm}{!}{\includegraphics{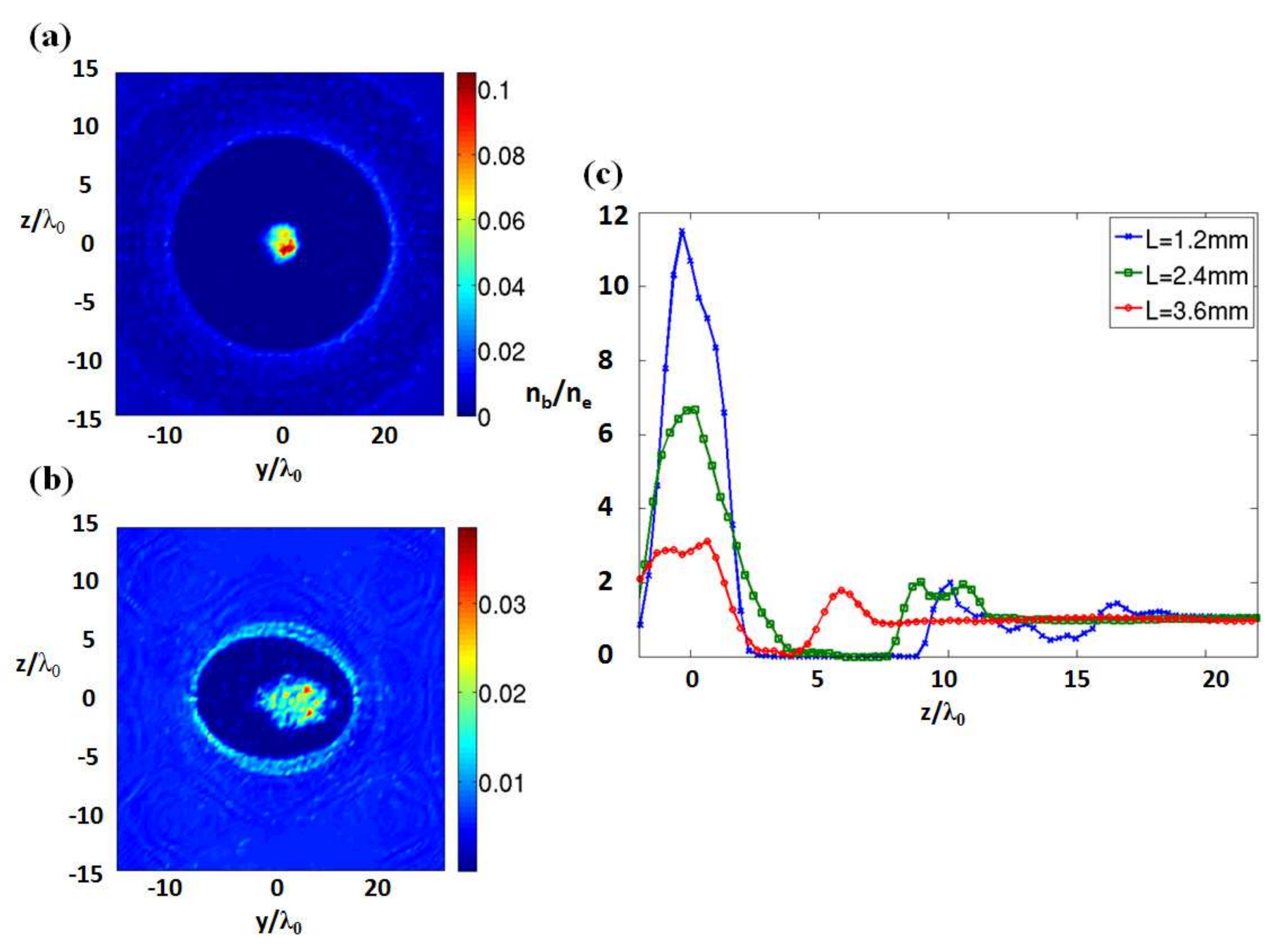}}
\caption{(a) and (b) Transverse cross-section of electron density, taken in the middle of the bubble in the longitudinal direction, in the LWFA regime (t=4 ps, at a propagation distance of L=1.2 mm) and in PWFA regime (t=12ps, at a propagation distance of L=3.6 mm). (c) line-out of the electron density (normalized to the background density $n_e$) along the transverse direction for different propagation distances $L=[1.2;2.4;3.6]mm$.}
 \label{Fig9}
\end{figure}

During the PWFA regime, as observed on Fig.\ref{Fig7}(b), the transverse shape of the bubble has changed and the radius around the leading bunch is not the same as the radius in the middle, which is a typical behavior of the PWFA regime.
In the blow-out regime of PWFA, the blow-out radius is no longer given by the matching with the laser driver $k_pR\simeq2\sqrt{a_0}$, but in this case it is given by \cite{Lu2} $k_pR_{th}\simeq2\sqrt{\frac{n_{b0}}{ne}}(k_p\sigma_r)\equiv2\sqrt{\Lambda}$, where $\Lambda$ is the normalized charge per unit length. Based on the values in Table \ref{Tab1}, we can estimate the theoretical values for the bubble radius in the PWFA regime for two different times: at t=8 ps (L=2.4 mm), we will find $R_{th}\simeq8.5\lambda_0$, and at t=12 ps (L=3.6 mm) $R_{th}\simeq6\lambda_0$. On Fig.\ref{Fig9}(a) and (b) is illustrated the transverse cross-section of the electron density for two different times: for t=4 ps in LWFA regime and for t=12 ps in PWFA regime, and we can see that despite the fact that the bubble is maintained, its radius decreases. In Fig.\ref{Fig9}(c) is illustrated a line-out of cross-section of the electron density (here normalized to background density) for different times corresponding to (i) the LWFA regime (L=1.2 mm), (ii) the beginning of the PWFA regime (L=2.4 mm) and (iii) later in time during PWFA regime (L=3.6 mm). The observed values of the bubble radius at these times in the LWFA and PWFA regimes are consistent with theoretical predictions, based on the amplitude $a_0$ for the first case, and the estimates given above for the second case depending on the radius and density of the leading bunch.

As can be seen in Fig.\ref{Fig8} and discussed previously, the laser at this time is fully depleted and is strongly distorted from its initial shape. It is modified by self-phase modulation, resulting in pulse steepening and pulse compression and is also subject to a frequency red-shifting. The density variations in an accelerating structure, in particular in the central region between the driving bunch and the beam load where electron density is decreasing provide a dynamic dielectric response resulting in the red-shifting of the laser light frequency \cite{Zhu 2013}. This longer wavelength can be seen in Fig.\ref{Fig8}(b) but appears to be separated from the front of the pulse. This is due to the bunch of electrons injected inside the bubble driven in the PWFA regime (Fig.\ref{Fig7}(a)). Indeed, this longer wavelength of the laser can be estimated to be around $\lambda_0^{'}\simeq3\lambda_0$, resulting in a critical density around $n_c^{'}\simeq2\times10^{20}cm^{-3}$, while the bunch of injected electrons inside the bubble as can be seen in Fig.\ref{Fig7}(a) may have a density around $n_e\simeq1.8\times10^{20}cm^{-3}$ ($0.1n_c$ defined at $\lambda_0=0.8\mu m$). Consequently, this bunch of electrons will be overcritical for the longer wavelength of the laser resulting in a {'}trapped{'} radiation pulse separated from the front of the pulse as can be seen in Fig.\ref{Fig8}(b).

\begin{figure}[!h]
\resizebox{13.cm}{!}{\includegraphics{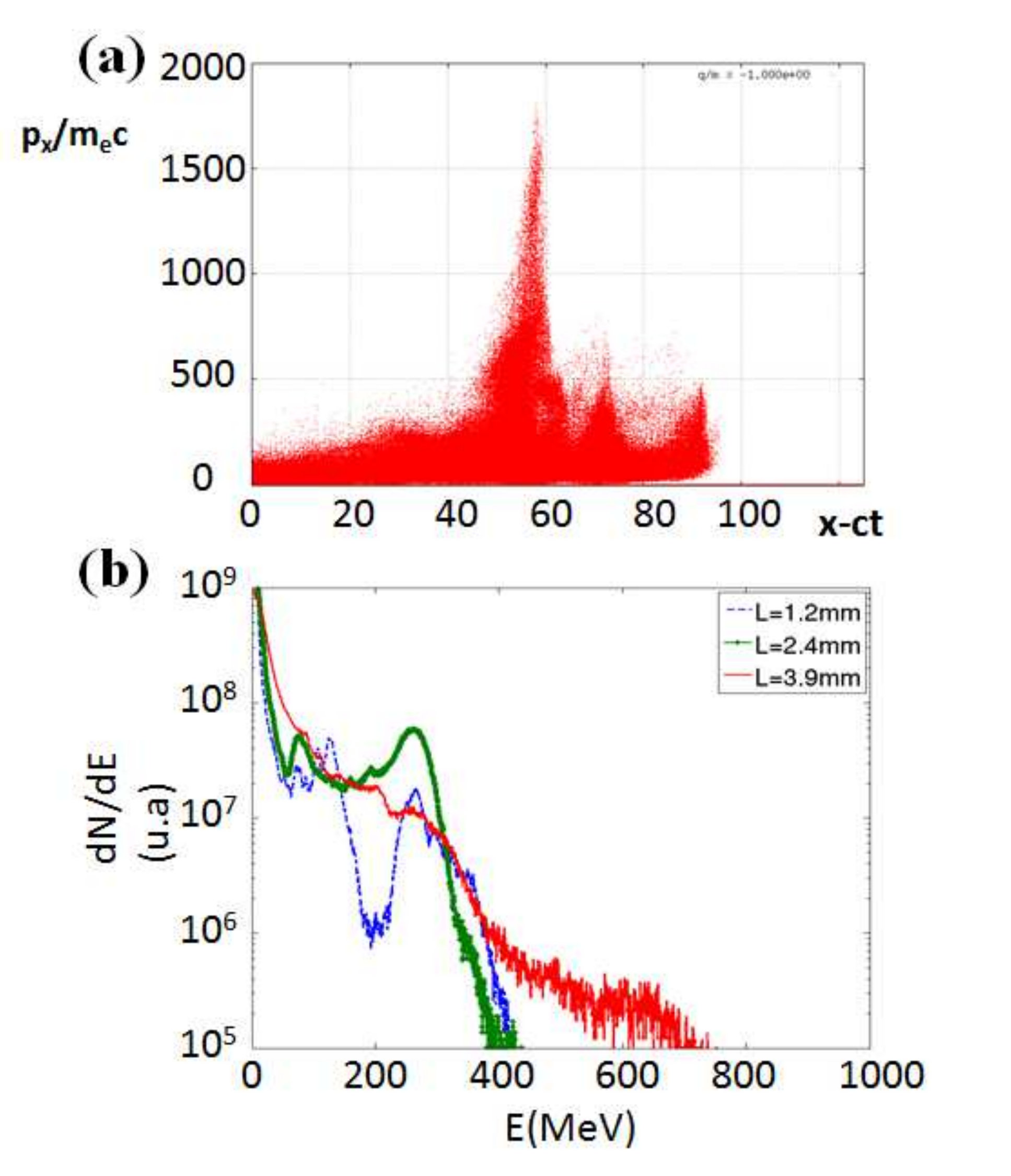}}
\caption{(a) Electron normalized momentum $p_x/mc$ after a propagation distance of L=4.3mm and (b) distribution function of electrons $dN/dE$ as a function of energy in MeV for three propagation distance corresponding to the LWFA regime (1.4mm), beginning of PWFA (2.4mm) and later in PWFA regime (3.9mm).}
 \label{Fig10}
\end{figure}

At late time, electrons that are accelerated will finally obtain $\gamma\simeq1500$ so almost 750MeV. As can be seen on Fig.\ref{Fig10}, showing the  distribution function $dN/dE$, the boosting effect due to the transition to PWFA is apparent. Initially, in the LWFA regime, the energy of the injected electrons can be estimated from the scaling law as shown in Eqn.\ref{LuScalingLaw}. This gives for our parameters a maximum electron energy around $E\simeq270MeV$, which is close to what we obtain in the LWFA regime (blue dashed curve in Fig.\ref{Fig10}).
Nevertheless, the boosting effect due to the transition to PWFA, while the laser is depleted, can clearly be observed and will accelerate some electrons up to 750 MeV, for a estimated charge around 8 pC for a peak energy of $700\pm100MeV$. In our simulations, with the resolution we were able to use, we have not been able to reproduce the experimental GeV level. To obtain a more realistic simulation of the experimental observations, a better resolution (in terms of number of particles) will be needed, because only a small number of electrons will be accelerated to this level. Also, small variations in the density of the plasma could change the initial laser propagation and consequently the injection/acceleration process during LWFA, and thus the evolution of the leading electron bunch. Changing density of the bunch will result in a different PWFA regime, and could increase the maximum energy obtained in the simulation. In other words, the  physical mechanism that can explain the high energies (larger than scaling due to LWFA in bubble regime) identified in our simulation, appears to be very sensitive to the characteristics of the first injected bunch during LWFA.

\begin{figure}[!h]
\resizebox{13.cm}{!}{\includegraphics{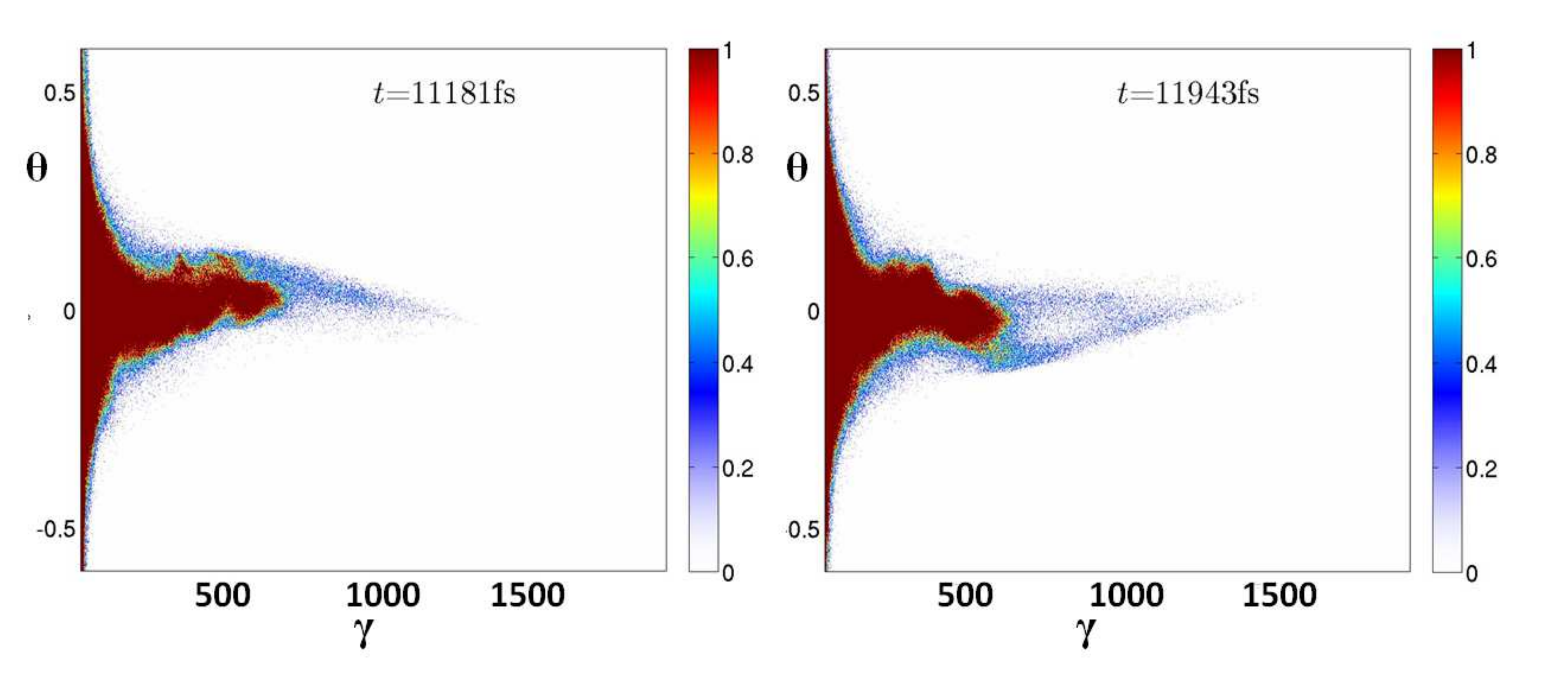}}
\caption{Normalized distribution of electrons as a function of energy ($\gamma$ factor), and the angle $\theta$ (in radians) defined as $\theta\equiv p_y/p_x$ for two different propagation distances L=3.3 mm (left) and L=3.5 mm showing the large transverse oscillations.}
 \label{Fig11}
\end{figure}

Finally, the PWFA stage finishes and the acceleration of electrons stops, when the leading bunch of electrons outrun the rest of the laser field. The electrons end up oscillating in the small, but still present laser field, and won't be able to sustain the bubble anymore, which stops the acceleration in the PWFA regime. Because of these oscillations in the laser field, the electrons from the leading bunch transfer this oscillatory motion to the entire bubble, which is oscillating transversely, as we can see in Fig.\ref{Fig11}, showing the normalized distribution $dN/dE$ as function of the angle $\theta$ defined as $\theta\equiv p_y/p_x$, and the $\gamma$ factor. We can only see oscillations of the bubble in the plane of polarization of the laser (here along $y$) and nothing can be seen in the $z$-plan, which clearly indicates that this is the oscillation of the leading bunch in the laser which creates these large oscillations of the entire cavity.
These oscillations of the accelerating cavity lead to acceleration of electrons off-axis, resulting in betatron oscillations inside the bubble of these accelerated electrons\cite{Kiselev,Rousse,Popp,Phuoc06,Albert2013,Schnell13}.

\section{Discussion}\label{sec4}

In this paper, we have presented experimental results of wakefield acceleration in 5 mm gas jet targets, showing the generation of GeV electrons in Helium.  This high level of energy, which is in disagreement with bubble wakefield scaling laws, can now be explained in view of the analysis using 3D PIC simulations by means of a two-stage process where an initial laser wakefield acceleration regime is followed by a plasma wakefield acceleration regime.  The experimental observations are consistent with a number of signatures of this two step process which can be derived both from the analytical scaling laws and from the 3D simulation results presented.

Firstly, the pump depletion length is expected to be of the order of 2mm, much less than the length of the gas jet.  At the densities of the current experiments, $10^{19}$ $cm^{-3}$ the depletion of the laser pump beam occurs quite rapidly and there is not enough laser power to drive the acceleration of the bubble across the full length of the gas jet.  Indeed this can be seen when looking at the Thomson scattered light from the interaction process which is only observed of over the first 1 to 2 mm of the interaction process as shown in Fig. \ref{Topview}.  In addition, the dephasing length for the LWFA process is similar, of the order of 2 mm, and thus even if the laser pulse were not depleting the electrons would start to dephase and lose energy at this point.  However, if the electron bunch has a large enough charge it will start to perturb the bubble field (beam loading) and as the laser pulse dies it will take over as the driver of the bubble.  During the transition from laser driven to electron driven bubble the perturbation in the bubble shape and plasma dynamics can potentially aid in the injection of a second electron bunch.

Once a secondary bunch of electrons is injected into the tail of the cavity it can then be accelerated throughout the remaining length of the gas jet system to energies which are much higher than those of the driving electron bunch as has already been observed in pure plasma wakefield accelerator experiments.  From the 3D PIC simulation it is observed that the second bunch reaches an energy of approximately double that of the driving electrons.  Because there is no longer a dephasing length limit the resultant energy could in principle be even greater than this, depending on the stength of the charge of the driving bunch and the length of interaction distance available.  In the simulations of Pae et al. \cite{Pae} it appears that in the case where the electron density was slightly lower, $7 \times 10^{18}$ $cm^{-3}$, the primary bunch of electron only had a charge of 200 pC and only led to acceleration of the secondary bunch up to 320 MeV, a similar energy to that of the primary bunch.   In the present simulation the larger charge in the primary bunch, of the order of 1nC, allowed for the much stronger acceleration of the secondary bunch up to double the driver electron energy.

As the primary electron bunch overtakes the laser pulse, the oscillatory EM field of the laser pulse is still sufficient to penetrate into the cavity and start to perturb the secondary acceleration process. The laser pulse at this point is strongly distorted due to self phase modulation and self steepening.  As can be seen in Fig.\ref{Fig8}(b) the leading edge is compressed while the tail of the pulse stretched to a lower frequency of approximately 1/3 of the initial frequency.  Due to this redshift, the radiation is now trapped within the tail region of the bubble because of its much lower critical density.  This residual field has the effect of inducing transverse oscillations in the bubble structure and in the secondary electron bunch which in turn could lead to betatron oscillations and enhanced betatron emission.  At the same time it perturbs the acceleration process and leads to a termination of the strong acceleration phase.

From the above, one would conclude that the combined LWFA/PWFA process requires a number of conditions in order to lead to effective enhancement of the electron energies above those obtained by pure LWFA in a similar system.  Firstly, a large charge bunch of primary electrons must be produced in order to drive the secondary PWFA process.  This requires good injection and fairly high electron density in order to create such a bunch.  It would be expected that this charge bunch should be of the order of charge that would cause significant beam loading effects since this bunch eventually should create a field strong enough to take over driving the wakefield bubble by itself.  Secondly, the pump depletion length should be approximately equal to the dephasing length so that just as the pump starts to fade away the primary electron bunch reaches the front of the bubble to start driving the wake itself.  These lengths should be significantly shorter than the gas jet length in order for the subsequent plasma wakefield process to be effective in accelerating a secondary bunch of electrons.  Both of these conditions would indicate that the observation of significant PWFA acceleration would require higher density plasmas setting a lower density limit for observing significant enhancements.  Indeed, looking at the experimental results shown in Fig.\ref{PEnery} it appears that the enhancements are seen for electron densities above approximately $8 \times 10^{19}$ $cm^{-3}$.  On the other hand as one goes to much higher densities the initial wakefield acceleration process no longer would produce distinct electron bunches and would lead to the heating of a broad distribution of electron energies instead, thus dispersing the primary electron bunch and reducing its effectiveness in driving the PWFA acceleration process. This is compounded by the fact that as the density increases, an even higher charge density will be required to effectively drive the process in the higher density plasma.  At the same time, at the higher densities the initial peak electron energy reduces with density and thus the boosted electron energies will also drop accordingly.  Thus, one might expect that the combined process becomes less effective at higher densities.  These two conditions lead to a window of densities where one could obtain the maximum boosted energies for a given laser power, wavelength and focal geometry.  In our case as can be seen from Fig.\ref{PEnery} it appears that this density range is approximately $8 \times 10^{18}$ $cm^{-3}$  to $2 \times 10^{19}$ $cm^{-3}$.

\section{Conclusion}\label{sec5}

In this paper, we have presented experimental results of wakefield acceleration, showing the generation of GeV electrons in Helium. These high electron energies, which are approximately double those from analytical laser bubble wakefield scaling laws can be understood as a laser wakefield process followed by a plasma wakefield process.  The characteristics of this two stage process are clearly identified in the 3D PIC simulations under conditions similar to those of the experiment.  The key components of the process include the creation of a large primary electron charge bunch, a pump depletion length approximately equal to the dephasing length so that the primary bunch can take over driving the plasma wake just as the laser pulse loses its driving strength and sufficient remaining plasma length for the plasma wakefield acceleration to boost a second bunch of electrons up to GeV energies.  These conditions can be met within a range of densities which in the case of the current experiment is approximately in the range of $8 \times 10^{18}$ $cm^{-3}$ to $2 \times 10^{19}$ $cm^{-3}$.  The present results indicate that attainment of energies approximately double those from LWFA alone can be achievable under well controlled conditions.  Clearly, further work is required both experimentally and theoretically to understand the detailed characteristics of this two stage process.

This work was funded by the Natural Sciences and Engineering Research Council of Canada and the Canada Research Chair program. One of the authors (P.-E. M.-L) would like to thank Denis Pesme, Stefan H\"{u}ller and Scott Wilks for many useful discussions. Two of the authors (M.Z. Mo and A. Atif) would like to thank the Canadian Institute for Photonic Innovations for providing travel funds. We thank ALLS technical team for their expert assistance throughout the experiments. The ALLS facility was funded by the Canadian Foundation for Innovation (CFI).

\end{document}